\documentclass[twocolumn,prd,aps,groupedaddress,showpacs,
nofootinbib,preprintnumbers]{revtex4}

\usepackage{amsmath}
\usepackage{amsfonts}
\usepackage{graphicx}
\usepackage{bm}

\def\be{\begin{equation}}
\def\ee{\end{equation}}
\def\ba{\begin{eqnarray}}
\def\ea{\end{eqnarray}}
\def\bs{\begin{subequations}}
\def\es{\end{subequations}}

\def\p{\partial}
\newcommand{\Eq}[1]{(\ref{#1})}

\begin{document}

\title{Stability of multi-field cosmological solutions}
\author{Gianluca Calcagni}
\email{g.calcagni@sussex.ac.uk}
\author{Andrew R. Liddle} 
\affiliation{Astronomy Centre, University of Sussex, Brighton BN1 9QH,
United Kingdom}
\date{November 21st, 2007}

\begin{abstract}
We explore the stability properties of multi-field solutions of
assisted inflation type, where several fields collectively evolve to
the same configuration. In the case of noninteracting fields, we show
that the condition for such solutions to be stable is less restrictive
than that required for tracking in quintessence models.  Our results,
which do not rely on the slow-roll approximation, further indicate
that to linear order in homogeneous perturbations the fields are in
fact unaware of each other's existence. We end by generalizing our
results to some cases of interacting fields and to other background
solutions and dynamics, including the high-energy braneworld.
\end{abstract}

\preprint{arXiv:0711.3360 [astro-ph]}

\maketitle


\section{Introduction}

The dynamical possibilities in multi-field inflationary scenarios are
considerably richer than in single-field models. An example is
assisted inflation \cite{LMS,MW}, where a collection of fields evolve
together, perhaps driving inflation that would not be possible if only
one field were present. Configurations with multiple scalar fields
have been considered in the case of exponential potentials with equal
or differing slopes. Later extensions of these models include the
effect of curvature and a barotropic perfect fluid \cite{CV}, Bianchi
backgrounds \cite{ACCZ}, and polynomial potentials \cite{KO}. The
stability of solutions as critical points in phase space was studied
for decoupled \cite{MW,GPZ,KLT} and coupled \cite{GPCZ,CNV,HPVW}
exponential potentials, and for decoupled inverse power-law potentials
\cite{KLT}. Late-time dark-energy scenarios can be found in
Refs.~\cite{KLT,Tsu06}. Multi-field scenarios naturally emerge in
Kaluza--Klein compactifications \cite{KO} and in M-theory, via the
nonperturbative dynamics of a stack of M5-branes distributed along a
$S^1/\mathbb{Z}_2$ orbifold \cite{BBK}. In the latter case, the scalar
fields and their exponential potentials are related, respectively, to
the distances and instanton interactions between the branes. A model motivated by string
theory is $N$-flation, where the scalar fields are axions \cite{DKMW}.

Thus far, general conditions for the existence of classically stable
assisted inflation solutions have not been given in the
literature.\footnote{Inhomogeneous perturbations of multi-field
configurations were considered in Ref.~\cite{HN}.} This is in contrast to
the related case of tracking quintessence models, featuring a single
field and a barotropic fluid, where the tracking condition was found
to be \cite{SWZ}
\be\label{trac}
\Gamma \equiv \frac{VV''}{V'^2} > 1 - \frac{2-\gamma}{4+2\gamma} \,,
\ee
where $V(\phi)$ is the scalar field potential, $\gamma$ the fluid
barotropic index defined by $p = (\gamma -1)\rho$, and prime a
derivative with respect to the field value $\phi$. The condition is
often simplified to $\Gamma > 1$, as otherwise the field energy
density does not grow relative to the fluid \cite{SWZ}.

The aim of this article is to derive those stability
conditions. In the first instance we are interested in the case of $n$
scalar fields $\phi_i$ each with an identical potential
$V(\phi_i)$. In that case Kim {\it et al.}~\cite{KLT} showed that,
provided the fields all have equal value, the system was dynamically
equivalent to a single-field model with field $\chi$ and potential
$W(\chi) \equiv n V(\chi/\sqrt{n})$. This generalizes the
correspondence for exponential potentials derived in Ref.~\cite{LMS},
and also shows there is no assisted inflation phenomenon for inverse
power-law potentials \cite{KLT}. However, while such solutions
obviously exist, that paper did not address under what circumstances
they are stable, which we do here. We then also consider some
generalizations of this basic set-up. None of our results require the
Universe to accelerate, so they are valid also beyond the slow-roll
approximation.

\section{Background evolution}

The background equations of motion with an Einstein--Hilbert
gravitational action and $n$ Klein--Gordon homogeneous scalars read 
\ba
&&H^2=\frac{\kappa^2}{3}
\left(W + \frac{1}{2} \sum_{i=0}^{n-1}
\dot\phi_i^2\right)\,,\\  
&&\ddot\phi_i+3H\dot\phi_i+\p_i W=0\,,\label{kge}
\ea
where $H$ is the Hubble parameter, $\kappa$ is the gravitational
coupling constant, dots represent derivatives with respect to
synchronous time $t$, $W\equiv
W(\phi_0,\phi_1,\dots,\phi_{n-1})$ is the potential with both
self-interaction and interaction terms, and $\p_i\equiv 
\p/\p\phi_i$. For simplicity, we set $\kappa^2=1$. The two equations can
be combined to give
\be\label{hj0}
\dot H =-\frac12\sum_i\dot\phi_i^2\,.
\ee

We are going to discuss the classical stability (that is, against
homogeneous perturbations) of solutions of the form 
\be\label{attr}
\phi_i(t)=\phi(t)\,\qquad i=0,\dots, n-1\,,
\ee
i.e.\ where the fields all evolve together. This class of solutions
includes multi-field fixed point solutions, e.g.\ as in
Refs.~\cite{CV,KLT}, but is much more general --- the solutions need
not be of scaling type (commonly defined as the scalar field kinetic
and potential energies maintaining a fixed ratio,
e.g.~Ref.~\cite{LS}), and indeed need not even be inflationary.

We begin by determining
what forms of potential support such solutions.
Denoting derivatives with respect to $\phi$ with primes, this solution
is possible only if 
\be\label{cond}
\p_iW\big|_{\phi_j=\phi}=V'(\phi)\qquad \forall i,j \,,
\ee
where in the left-hand side $\phi_j=\phi$ for all $j=0,\dots n-1$, and
$V'$ is defined by this equation. 
If the fields are mutually decoupled, then each can be written with
its own potential $V_i$ obeying 
\be\label{V}
V_i(x)+\Lambda_i=V(x)
\ee
for all $i$. Here the $\Lambda_i$ are constants, which can all be
absorbed into some $\Lambda=\sum_i \Lambda_i$ acting as a cosmological
constant, so that $W = \sum_i V_i(\phi_i)+\Lambda$. We will see that
this constant does not affect whether the solution is stable or not,
though it can alter the qualitative behaviour in the vicinity of the
solution. Ordinarily
we will consider $\Lambda$ to be zero or negligibly 
small, so that all the fields have the same potential.

If one allows for a coupling between the fields, it must still satisfy
Eq.~\Eq{cond}. This happens, for instance, when the interaction term
is symmetric in all the fields, so that one can write the potential as
\be\label{sym}
W=\sum_i V_i(\phi_i)+f\left[\prod_k g(\phi_k)\right]\,,
\ee
for some functions $f$ and $g$.  Also, there can be more general
situations, as the following three-field system shows:
$W=\phi_0^a+a(2b-a)^{-1}\phi_0^{a-b}\phi_1^b+b(2b-a)^{-1}\phi_2^a$.

\section{Stability}

Let us first consider the simple case of an expanding universe ($H>0$)
and no interaction terms, $V_i=V$. We do so using two different
formalisms, the latter of which will prove the more powerful.

\subsection{Synchronous time formalism}

Perturbing the above equations of motion around the solution
Eq.~\Eq{attr}, 
\be
H(t)\to H(t)+\delta H(t)\,\qquad \phi_i(t)\to \phi(t)+\delta\phi_i(t)\,,
\ee
and defining the $(2n+1)$-vector ($T$ denotes transposition)
\be
X\equiv (H,\phi_i,\dot\phi_i)^T\,,
\ee
the linearized equations of motion can be written in a matrix form
$\dot{\delta X}=M\delta X$, where the entries $m_{ij}=(M)_{ij}$ are 
\ba
&&m_{ij}=0\,\qquad i,j=0,\dots, n\,,\nonumber\\
&&m_{0j}=-\dot\phi\,,\qquad j=n+1,\dots, 2n\,,\nonumber\\
&&m_{ij}=\delta_{ij}\,,\qquad j=i+n=n+1,\dots,2n\nonumber\\
&&m_{i0}=-3\dot\phi\,,\qquad i=n+1,\dots, 2n\,,\nonumber\\
&&m_{ij}=-V''\delta_{ij}\,,\qquad i=j+n=n+1,\dots, 2n\,,\nonumber\\
&&m_{ij}=-3H\delta_{ij}\,,\qquad i,j=n+1,\dots, 2n\,,
\ea
where
\be
V''(\phi) \equiv \p_i^2 W\big|_{\phi_j=\phi}\,.
\ee
The characteristic equation $\det(M-\lambda I_{2n+1})=0$ determines
the eigenvalues $\lambda=0$ and 
\ba
\lambda_\pm^*
&=&-\frac{3H}{2}\left(1\pm\sqrt{1-\frac{4}{9}\frac{V''}{H^2}}\right)
\,\label{eige},\\  
\tilde\lambda_\pm
&=&-\frac{3H}{2}\left[1\pm\sqrt{1-\frac{4(V''-3n\dot\phi^2)}{9H^2}} 
   \right]\,,\label{eigs} 
\ea
where $\lambda_\pm^*$ is $n-1$ times degenerate. 

The solution is stable provided no eigenvalue has a positive real part.
Equation \Eq{eige} imposes the simple stability condition 
\be\label{stabi}
V''\geq 0\,,
\ee
while the real part of $\tilde\lambda_-$ is negative
definite only if a stronger condition is valid: 
\be\label{track}
V''\geq 3n\dot\phi^2\,.
\ee
In the slow-roll approximation $3H^2\approx nV$,
$-3H\dot\phi\approx 
V'$, this is nothing but the tracking condition 
\be\label{track2}
\Gamma\equiv\frac{VV''}{V'^2}\geq 1\,,
\ee
but note that Eq.~\Eq{track} is valid also outside inflation.

This looks like a natural conclusion exactly matching the expectation
inherited from the known quintessence case. However in fact the
interpretation is quite different, as can be seen by considering the
implications in the simple situation of the power-law potential
\be
V = \phi^p\,,\qquad \Gamma=1-\frac1p\,.
\ee
Clearly, this potential with $p>0$ (in particular, $p=2$) does not
satisfy the condition \Eq{track}, implying Eq.~\Eq{attr} is not a
slow-roll attractor, contradicting the known result \cite{BW} that
solutions for this potential do approach radial trajectories which, by
field rotation, can be described as all the fields rolling together
(see also Ref.~\cite{UR}).

The reason for this discrepancy is that this formalism fails to
recognize that trajectories distinguished only by a shift in time,
$t\to t+\delta t$, are physically equivalent and should be identified,
even if the apparent perturbation \emph{at fixed time} appears to
grow. This differs from the case of a tracking quintessence model,
where the existence of a barotropic fluid acts as a ``clock'', with
reference to which time shifts in the scalar field evolution can be
physically distinguished.

The time-shift mode can be removed by defining quantities which are
gauge-invariant with respect to the gauge group of gravitation (group
of infinitesimal coordinate transformations) \cite{MFB}. In the
particular case of multiple scalar fields and homogeneous
perturbations, this is
$\delta\tilde\phi_i=\delta\phi_i(t)-\dot\phi\delta t$
\cite{HN}. Rather than writing down gauge-ready equations using the
synchronous time formalism, we now employ a simpler alternative.

\subsection{Hamilton--Jacobi formalism}

In the Hamilton--Jacobi (HJ) formalism \cite{Mus90,SB}, the role of
the clock is invested in a matter field, so that one can get rid of
the unphysical degree of freedom represented by shifts in the time
coordinate $t$. When many matter fields are present, one can choose
one and study homogeneous perturbations on constant time slices. More
specifically, we can take $\phi_0=\phi$ as the new time coordinate and
perturb only the other $n-1$ fields $\phi_i(\phi)$. The equation of
motion for $\phi$ is a constraint determining the relation between
different time coordinates. The first HJ equation is given by
Eq.~\Eq{hj0}, 
\be\label{hj1} 
\dot\phi=-\frac{2H'(\phi)}{A}\,, 
\ee
where 
\be 
A\equiv 1+\sum_i\phi_i'^2\,; 
\ee 
here and in what follows $i$ runs from 1 to $n-1$.

Since $\p_t^2=\dot\phi^2 \p_\phi^2
-(3H\dot\phi+W')\p_\phi$, one gets $n-1$ equations of the form
\be\label{hj2}
\dot\phi^2\phi_i''-W'\phi_i'+\p_iW=0\,.
\ee
The expansion of the universe is encoded in the friction term and in
$\dot\phi^2$ via Eq.~\Eq{hj1}. 
Plugging Eq.~\Eq{hj1} in the Friedmann equation one gets
\be\label{hj3}
2H'^2-3A H^2+AW=0\,.
\ee
When perturbing Eqs.~\Eq{hj2} and \Eq{hj3} around Eq.~\Eq{attr}, some
simplifications occur, since $\phi_i''=0$ and $A=n$ on the background
($\delta A=2\sum_i\delta\phi_i'$). The final result is
\ba
&& \dot\phi^2\delta\phi_i''-V'\delta\phi_i' +\sum_j
(\p_j\p_iW-\p_j W')\delta\phi_j=0\,,\label{dhj1}\\ 
&& \frac{4}{n}H'\delta H'-6H\delta
H+\sum_i\left(V'\delta\phi_i-\dot\phi^2\delta\phi_i'\right) =
0\,, \quad \label{dhj2}  
\ea
where all background quantities are functions of $\phi$. In general,
the matrix $\p_j\p_iW$ is symmetric with all different entries.

When the fields are decoupled, $\p_j\p_iW(\phi)=V''(\phi)\delta_{ij}$,
$\p_j W'=0$, 
and $j=i$ in Eq.~\Eq{dhj1}. For simplicity we shall concentrate on
this case. Again, the system can be written as $\delta X'=M\delta X$,
where $X$ is now a $(2n-1)$-vector and the coefficients of the matrix
$M$ are given by Eqs.~\Eq{dhj1} and \Eq{dhj2}. Define the objects
\ba
\lambda_0 &=& -\frac{3H}{\dot\phi}\,,\label{eq25}\\
\lambda_\pm &=&
\frac{V'}{2\dot\phi^2}\left(1\pm\sqrt{1-\frac{4\dot\phi^2
      V''}{V'^2}}\, \right)\,. \label{hjeig}
\ea
The eigenvalues of the decoupled system are $\lambda_0$ and the
$(n-1)$-times degenerate pair $\lambda_\pm$. Concerning the first
eigenvalue, it is normally convenient to take $\dot{\phi}>0$ (arranged
if necessary by sending $\phi \rightarrow -\phi$) so that increasing
$\phi$ corresponds to increasing time. The first eigenvalue is then
negative definite, as required for stability.  If $\dot{\phi}$ were
negative, then the two time choices would flow in opposite directions
and the stability condition would become a positive-definite
eigenvalue, the physics being unchanged. Therefore Eq.~(\ref{eq25})
holds in either case.

With the convention $\dot\phi>0$ one has that $V'<0$, as the
situations of physical interest always have the field rolling down the
potential and hence $\dot\phi$ should have the opposite sign to $V'$.
If $V''>0$ [Eq.~\Eq{stabi}] holds there are two decaying modes related to
$\lambda_\pm$ (with the equality, one direction is flat) and
Eq.~\Eq{attr} is a classical attractor. If $V''>V'^2/4\dot\phi^2$,
the perturbations periodically overshoot the attractor.

Notice that
\be\label{exa}
\dot\phi^2\lambda_+\lambda_-=\lambda_+^*\lambda_-^*\,,
\ee
meaning that the perturbations of the $i$-th field ($i\neq 0$) are
related in both formalisms. However, they are not equal except in the
slow-roll regime. To show this, let us denote as $\delta\phi_i^{(t)}$
the perturbation of the $i$-th field in the $t$-formalism. Using the
diagonalized linear equations and Eq.~\Eq{exa}, one finds 
\be
\frac{\delta\ddot\phi_i^{(t)}}{\delta\phi_i^{(t)}} -
\frac{\delta\ddot\phi_i}{\delta\phi_i}  
= -\frac{\ddot\phi}{\dot\phi}\frac{\delta\dot\phi_i}{\delta\phi_i}\,.
\ee
The right-hand side is nonvanishing because $\delta\phi_i^{(t)}$ is
not gauge invariant. However, it is negligible in the slow-roll
approximation. This is expected since, in this case, the eigenvalues
would be almost constant and perturbations would be exponential:
$\delta\phi_i\sim e^{\lambda_\pm^* t}$ in the $t$-formalism, while
$\delta\phi_i\sim e^{\lambda_\pm \phi}$ in the HJ formalism. Then the
eigenvalues would be individually related, $\lambda_\pm^*\approx
\dot\phi\lambda_\pm$.

\subsection{Interpretation}

The conclusion from these calculations is that the stability condition
for multi-field tracking in identical potentials is simply $V''>0$,
rather than a condition similar to the quintessence tracking
condition. The interpretation is in fact straightforward, and most
easily seen by studying Eq.~(\ref{dhj1}) for the perturbed scalar
field evolution. This equation is \emph{independent} of the perturbed
metric $\delta H$. Accordingly, a particular field perturbation is
insensitive, at linear order, to the perturbation in the metric caused
by any other field perturbation. The fields evolve independently of
each other. Whether or not the field perturbation dies away, i.e.~the
fields approach one another, simply depends on whether those fields
which happen to be further down the potential are evolving more
slowly, which is the case provided the potential is less steep
there. Assisted inflation does take place, but not because the fields
are actually aware of each other's value and are drawn towards one
another, but rather just as a generic property of trajectories on
convex potentials.

Multi-field systems which do not satisfy Eq.~\Eq{stabi} will never
attain the configuration Eq.~\Eq{attr}. On the other hand,
Eq.~\Eq{stabi} guarantees the stability of the equal-field solution
Eq.~\Eq{attr} to small perturbations, but does not tell us whether
such a solution acts as a \emph{global} attractor for the system. In
the simple case of shallow exponential potentials, this global
attractor property is known to hold \cite{CV}. However, it is easy to
find counterexamples. Consider the case of steep exponential
potentials $V_i(\phi_i) \propto \exp(\lambda \phi_i)$ for
$\lambda^2>6$. These potentials are so steep that there are no scaling
solutions, as the potential energy falls off faster than the kinetic
energy. Our particular solution is stable to small perturbations
($V''=\lambda^2 V>0$), but if the fields are initially set well apart
the rapid decay of their velocity will prevent them asymptoting to
equal field values.

Note also that our condition $V''>0$ is a local stability condition
referring to the position on the potential that the fields happen to
be at during a given epoch. Some potentials may have $V''$ with the
same sign everywhere, and the stability or otherwise is then a global
property of the potential. But potentials where $V''$ changes sign may
experience sequences of epochs where the fields alternately converge
or diverge from each other, the ultimate stability being determined by
the sign of $V''$ after its final sign change.

We mention briefly the effect of lifting the assumption $\Lambda = 0$,
which changes the value of $H$ corresponding to a given location of
the scalar fields (we assume we stay on the expanding branch). Looking
at the eigenvalue equation Eq.~\Eq{eige}, we see that changing the
value of $H$ can modify (or create) an imaginary part to the
eigenvalues but can change the real part only by an overall
multiplier, leaving its sign intact. Hence introducing $\Lambda$ can
change the type of attractor (oscillatory or monotonic), but not
whether the potential gives stable solutions or not. A positive
$\Lambda$ can also impact on whether or not the solution Eq.~\Eq{attr}
is a global attractor, in the sense specified above. Such cosmologies
will be asymptotically de Sitter, with rapidly-decaying field
velocities that may prevent approach to the equal-field solution.

The impact of $\Lambda$ is somewhat less transparent in the
Hamilton--Jacobi formalism, Eq.~\Eq{hjeig}, where at first sight
$\lambda_\pm$ seems independent of $H$ even though we have verified
that within slow-roll $\lambda_\pm^* \approx \dot{\phi}
\lambda_\pm$. The resolution is that at a given location on the
potential, the introduction of $\Lambda$ will modify the value of
$\dot{\phi}$ in Eq.~\Eq{hjeig}, again potentially introducing or
modifying an imaginary part to the eigenvalue.

One might ask how our conclusions would change if a fluid component
with equation of state $p=(\gamma-1)\rho$ were added, to give a
multi-field quintessence scenario. Adding a new component changes the
nature of the dynamical system, and the new degree of freedom may well
induce an instability, even if the fluid component is taken as
initially sub-dominant. E.g., for sufficiently-steep exponential
potentials the late-time global attractor has the field scaling with
the fluid \cite{CV}. Nevertheless, one would expect that within the
field sector the stability condition remains the same, i.e.\ provided
$V''>0$, the stable solution will still maintain equal field
values. It is easy to see that this is true if $\rho=0$ in the
background solution, provided $\gamma>0$ (the extra degree of freedom
would decouple from the scalar fields with eigenvalue $\propto -\gamma
H/\dot\phi$).

Notice also that when $\gamma=2$ (stiff matter, $p=\rho$)
Eq.~\Eq{trac} reduces to the pseudo-tracking condition Eq.~\Eq{track2}
found in synchronous time for slowly-rolling fields. Stiff matter
decays as $\rho \sim a^{-6}$, this being the fastest rate at which a
scalar field density can be diluted (kinetic regime, $\dot\phi^2\gg
V$). Since slow rolling prevents this condition, for all purposes the
fluid is negligible relative to the scalars, which is precisely the
configuration leading to Eqs.~\Eq{track} and \Eq{track2}. However,
this implies that one cannot use the barotropic fluid as a reliable
clock throughout the whole evolution of the system, and the
calculation in synchronous time becomes nonphysical. In other words,
classical stability in multi-field cosmologies with a barotropic fluid
can be consistently studied in synchronous time for $\gamma<2$ but not
in the limit $\gamma\to 2$ (no fluid), where the HJ formalism is more
appropriate.

\section{Generalizations}

\subsection{With field couplings}

The next simplest case is that of a symmetric interaction,
Eq.~\Eq{sym}, so that the nondiagonal entries of $\p_j\p_iW$ are all
equal, and we can define the useful quantity
\be
\label{y} y(\phi)\equiv\p_j\p_i W=\p_j\p_i f\,,\qquad j\neq i\,.
\ee
Defining
\be
\lambda_\pm^{(n)} =
\frac{V'}{2\dot\phi^2}\left[1\pm\sqrt{1-4\dot\phi^2\frac{
      V''+(n-2)y}{V'^2}}\,\right]\,,
\ee
one can show that the HJ eigenvalues are $\lambda_0$ and the
$(n-1)$-times degenerate pair $\lambda_\pm^{(1)}$. Stability is
achieved if  
\be\label{condy1}
V''\geq y\,.
\ee
We can also consider the case where the coupling is symmetric in all
the fields except one which remains uncoupled
($\p_j W'=0$). It is natural to choose the decoupled field as the
clock field $\phi=\phi_0$ (though one would be free to choose another). Then the 
eigenvalues are $\lambda_0$, $\lambda_\pm^{(n)}$, and the
$(n-2)$-times degenerate pair $\lambda_\pm^{(1)}$, while the stability
condition is
\be\label{condy}
V''\geq \max \left\{y,\, -(n-2)y\right\}\,.
\ee

With interactions included, stability now requires steeper self-interaction potentials, which is a
confirmation of the claim made in Refs.~\cite{ACCZ,KO} that multi-field
inflation with cross-couplings is less likely to happen. If $y<0$, the
larger the number of fields the steeper the coupling term.
As an application of Eq.~\Eq{condy}, consider $n$ scalars with 
potentials $V_0=C\phi^p$, $V_i=\phi_i^p$, and power-law interaction 
$f=B(\phi_1\cdots\phi_{n-1})^{p/(n-1)}$, where $B$, $p$ and 
$C=1+B/(n-1)$ are constants. Let also $B,p>0$. Then
$y=Bp^2\phi^{p-2}/(n-1)^2$ and the solution is stable if $p\geq C$. 

If the cross-coupling is asymmetric, the symmetric matrix
$\p_j\p_iW(\phi)$ and the eigenvalues of the linearized equation are
more complicated; however, for a given model it is straightforward to check
its classical stability via Eqs.~\Eq{dhj1} and \Eq{dhj2}.


\subsection{More general backgrounds and the braneworld}

We can generalize all the above results to other
background equations and solutions. It is instructive to consider
background solutions of the form 
\be\label{genbs}
\phi_i(t) = \alpha_i\phi^{\beta_i}(t)\,,
\ee
for some real $\alpha_i$ and $\beta_i$. Equation \Eq{genbs} includes
the solution Eq.~\Eq{track} ($\alpha_i=1=\beta_i$), scaling solutions
\cite{LMS,MW}, the power-law inflationary attractor studied in
Ref.~\cite{KLT}, and other situations where, for instance, some fields
are subdominant with respect to others ($\alpha_i\approx 0$ for some
$i$). The condition on the potentials for Eq.~\Eq{genbs} to be a
solution is 
\be\label{newW}
\p_iW\big|_{\phi_j=\phi} = \alpha_i\beta_i\phi^{\beta_i-1}
\left[V'(\phi)+(1-\beta_i)\frac{\dot\phi^2}{\phi}\right]\,.
\ee
Again, $V'$ is defined to be the potential of $\phi_0$, differentiated
with respect to $\phi_0$, where all fields are set to be equal to
$\phi_0$; in other words, Eq.~(\ref{cond}) with $i=0$.  This condition
is fulfilled, e.g., by slowly-rolling fields with different
polynomial potentials. On this background, $A(\phi) =
1+\sum_j(\alpha_j\beta_j)^2\phi^{2(\beta_j-1)}$ in Eq.~\Eq{hj1}.

In high-energy braneworld scenarios inflation can be sustained by
steeper potentials \cite{CLL} and it would be interesting to see how
the stability conditions are affected. Neglecting brane--bulk energy
exchange and the contribution of the Weyl 
tensor, the Klein--Gordon equation \Eq{kge} [as well as Eq.~\Eq{newW}]
is unchanged, while the effective Friedmann equation on the brane at
high energies reads 
\be\label{bFRW}
H^{2-\theta}=\frac{\kappa_\theta^2}{3} \rho\,.
\ee
Here, $\theta=0$ in the pure 4D regime, $\theta=1$ in the high-energy
Randall--Sundrum braneworld, and $\theta=-1$ in the high-energy
Gauss--Bonnet scenario; $\kappa_\theta$ is the effective gravitational
coupling felt by the matter (with energy density $\rho$) on the
brane. See Refs.~\cite{cal3} for a review of this formalism and
references to the original results. 

We continue to work with the more general background solution given by Eq.~\Eq{genbs}.
In units $\kappa_\theta^2=1-\theta/2$, the HJ equations become
\ba
\dot\phi &=& -\frac{2H'}{H^{\theta}A}\,,\label{bH}\\
H'^2 &=& \frac{AH^{2\theta}}{2}\left[
\frac{6H^{2-\theta}}{2-\theta}-W\right]\,.\label{bhj3}
\ea
Linearizing Eq.~\Eq{bhj3} one gets 
\ba\label{lbh}
\delta H'&=&
-\frac{H}{\dot\phi}(3+\theta\epsilon)\,\delta
H\nonumber\\
&&\quad+\frac{H^{\theta}}{4\dot\phi}\Big(2\sum_i\p_iW\delta\phi_i
-\dot\phi^2\delta A\Big)\,, 
\ea
with
\be
\epsilon\equiv -\frac{\dot H}{H^2}=\frac{A\dot\phi^2}{2H^{2-\theta}}\,,
\ee
and
\be
\delta A = 2\sum_i\alpha_i\beta_i\phi^{\beta_i-1}\delta\phi_i'\,.
\ee
Also, since $\phi_i'' =
\alpha_i\beta_i(\beta_i-1)\phi^{\beta_i-2}\neq 
0$, the perturbed Eq.~\Eq{hj2} now contains the metric perturbation: 
\ba
\delta\phi_i''&=&
-\phi_i''\frac{12H^{1-\theta}}{A\dot\phi^2}\,\delta H
+\sum_j\left(\frac{V'}{\dot\phi^2}\delta_{ij} +
\frac{2\phi''_i\phi'_j}{A}\right)\delta\phi'_j
\nonumber\\
&&\quad+\frac{1}{\dot\phi^2}\sum_j\left(
\frac{2\phi_i''}{A}\,\p_jW-\p_j\p_iW+\phi_i'\p_jW'\right)
\delta\phi_j\,.\nonumber\\\label{lbh2} 
\ea

The eigenvalue equations can be solved numerically or analytically for particular
cases. A simple possibility is $\beta_i=1$ for all $i$, whereupon $\phi_i''=0$ and many of the terms in Eq.~\Eq{lbh2} vanish. The background
condition on the total potential becomes $\p_iW=\alpha_i V'$, and the
perturbation of the Hubble parameter decouples from the scalar
ones. In this case, the form of the effective Friedmann equation
affects only Eq.~\Eq{lbh}. The eigenvalue $\lambda_0$ is positive
definite only if [see Eq.~\Eq{lbh}]
\be
-\theta\epsilon\leq 3\,,
\ee
which is automatically satisfied for $\theta\geq 0$ (4D and
Randall--Sundrum braneworld). In the Gauss--Bonnet scenario, it must be
$\epsilon\leq 3$, which is true during accelerating expansion. 

When there is no coupling between fields ($\p_j\p_iW=0$), the
other stability condition Eq.~\Eq{stabi} becomes 
\be
\alpha_i V''\geq 0\,.
\ee
For quadratic potentials this states that none of the scalars is
tachyonic. In the presence of interactions and for $\alpha_i\neq
\alpha_j$, the degeneracy of the eigenvalues is broken and more
involved relations arise.

\section{Conclusions}

We have derived a set of stability conditions against classical
perturbations for multi-field cosmological solutions.
These can be summarized as follows. 
\begin{itemize}
\item For the solution Eq.~\Eq{attr}, if the fields are noninteracting
  the only condition is $V''\geq 0$, while for symmetric interactions
  with positive coupling the self-interacting potentials must be
  steeper. 
\item For the solution Eq.~\Eq{genbs} with $\beta_i=1$ and no
cross-coupling, the stability condition reads $\alpha_i V''\geq 0$ for
all $i$.
\item These results are valid for both four-dimensional cosmology and
high-energy Randall--Sundrum braneworld. On a Gauss--Bonnet
braneworld, the extra condition $\epsilon\leq 3$ is required.
\item In all other cases, the linearized dynamical system can be
studied numerically via Eqs.~\Eq{lbh} and \Eq{lbh2}.
\end{itemize}

\acknowledgments
The work of G.C.\ is supported by a Marie Curie Intra-European
Fellowship under contract MEIF-CT-2006-024523. We thank Ed Copeland
and Paul Saffin for useful discussions.


\end{document}